\documentstyle[prb,twocolumn,aps]{revtex}

\begin{document} \tolerance 50000

\draft

\title{A Rotating-Valence-Bond scenario for the 2D Antiferromagnetic
Heisenberg Model} 
\author{Germ\'an Sierra$^{1}$\cite{ger} and
 Miguel A. Mart\'{\i}n-Delgado$^{2}$ } 
\address{ $^{1}$Theoretische Physik - ETH-Honggerberg CH-8093, Switzerland 
\\ 
$^{2}$Departamento de
F\'{\i}sica Te\'orica I, Universidad Complutense.  28040-Madrid, Spain
 \\
}

\twocolumn[ 

\date{January 96} \maketitle \widetext

\vspace*{-1.0truecm}

\begin{abstract} \begin{center} \parbox{14cm}{ We propose that the valence bonds forming the ground state of
the 2D-AF Heisenberg model on a square lattice
 may rotate under the effect of the
antiferromagnetic background. To test this idea we apply a real
space renormalization group approach to construct a
rotating-valence-bond ground state anstaz. Our results are analytic
and valid for any value of the spin S of the AF-magnet, which
allow us to perfom semiclassical expansions of the energy and the
staggered magnetization.} \end{center}
\end{abstract}

\pacs{ \hspace{1.9cm} PACS numbers: 05.50.+q, 75.10.-b, 75.10.Jm}
 ]


Despite of some initial controversies there is by now sufficient
theoretical and experimental evidence for the existence of
antiferromagnetic long range order 
(AF LRO) in the 2d spin 1/2 Heisenberg antiferromagnet
\cite{manousakis} (and references therein). 
This property has been observed in parent
compounds of hight-${T_c}$ materials such as $ La_2 Cu O_4$
\cite{manousakis}. 
From a theoretical point of view this  means that the
strong quantum fluctuations implied by the low dimensionality
and the spin 1/2 do not destroy completely the Neel order, as it
happens \cite{bethe} in 1d. Though there is no  a satisfactory
physical explanation of this fact, which may be important regarding
 the interplay between  antiferromagnetism and superconductivity
upon doping. The RVB scenario originally proposed by Anderson
\cite{anderson,kivelson}, 
while yielding an appealing picture of the ground
state, does not explain the  presence of AF LRO. This type of
order may however be incorporated a posteriori in long range RVB
ansatzs of factorized form  \cite{liang-doucot-anderson}, 
with predictions
similar to the ones obtained using Quantum Monte Carlo methods
\cite{carlson} and variational plus Lanczos techniques 
\cite{hebb-rice}. 
A class of physical systems  where the RVB approach may
be actually realized is in spin ladders with an even number of
chains \cite{white-noack-scalapino,dagotto-rice}. The previous works
leave still room to investigate in more depth the interplay between
the RVB scenario, or more generally  ``valence bond scenarios'',
and the AF order present in the 2d AF-magnets, described by the
AF Hamiltonian 
$H=J \sum _{\langle i,j \rangle} {\bf S}_i \cdot {\bf S}_j$.

In this letter we shall study a new  possible effect of the
antiferromagnetism on the dynamics of bonds,  namely
bond-rotation. Let us recall that the basic mechanism considered
by the RVB picture is the resonance between parallel neighbour
bonds, which causes a substantial reduction of the ground state
energy \cite{anderson-fazekas}. In addition to this effect we shall
explore the possibility that the bonds also rotate around their
ends under the influence of the AF background. To test this idea we
propose a variational  ground state in which the bonds rotate
but do not resonate among themselves. This simplified ansatz 
allows us to decouple the rotation of bonds  of other phenomena and
in particular the  resonance. A  more realistic ansatz would
include both  rotation and  resonance of bonds, but we shall not
deal with  this issue here. The analysis of the pure
rotational-valence-bond ansatz leads us to a value of the
staggered magnetization closer to the exact numerical result of
the AFH model  than to the Neel magnetization while  we find the opposite
result for the ground state energy. We interpret these results
in terms of the  previous picture of  a bond scenario compatible
with  AF LRO and not containing the resonating mechanism, which
is responsable for the lowering of the energy. An important
ingredient of our construction is the use of  real space
renormalization group techniques, which allows us to obtain exact
analytical results for any value of the spin $S$ of the model ($S$
is integer or half-integer and in the discussion above $S=1/2$). 
The physical reason for this is that every cluster of 5
spins which share a rotating bond,  behave as an effective spin
$3S$, coupled with its neighbours through a Heisenberg
interaction with an effective coupling constant. Hence the
effective spin renormalizes to infinity, which allows us to compare
the   staggered magnetization and  energy of the ansatz with the
semiclassical spin-wave $1/S$ expansions of these quantities for
the ground state of the AF-magnet. This comparison confirms the
above picture of the rotating-valence-bond state as a state
close to the exact ground state of the AF Heisenberg model in
staggered  magnetization  but not in energy.

Let us begin our approach by considering the cluster of 5 spins 1/2
of Fig.\ref{fig1} a). The configuration showed in Fig.\ref{fig1} a)  
is the exact
ground state of the Ising piece  of the Heisenberg Hamiltonian,
given by $H_{z} = J\sum_{i=1,\dots,4} S^z_0 S^z_i $ , where $S_0^z $
and $S_i^z$ are the third component of the spin operators at the
center and the $i^{th}$ position off the center respectively. 
As soon as the ``transverse'' Hamiltonian
 $H_{xy} = J \sum_{i=1,\dots,4} S^x_0 S^x + S^y_0 S^y_i $ is switched
on, the down-spin in the middle starts to move around the
cluster, and a valence bond between the center and the remaining
sites is formed in a s-wave ($l=0$) symmetric state as 
shown in Fig.\ref{fig1} b). Other rotational states with $l\neq 0$ may appear
corresponding to excitations ($l$ being the orbital angula momentum 
of the bond).
An alternative
description of this state is given by first combining the 4
spins sourrounding the center into a spin 2 irrep, which in turn is
combined with the spin 1/2 at the center yielding a spin 3/2 irrep
with energy $e_0 = - 3 J/2$. If instead of the spin 1/2 at
each site there is a spin S the previous analysis can be easily
generalized as follows: the ground state of the AFH Hamiltonian of
the 5-cluster has total spin 3S and is obtained by first combining
all the surrounding spins into a spin 4S, which in  turn
becomes a spin 3S after multiplication with the spin S at the
center. In a certain sense this state can be viewed as the formation
of bonds between the center and its four neighbours.
After applying several steps of the real-space RG, as we shall see 
below, new bonds are generated between sites at longer distances
apart. Thus our valence-bond scenario is a type of long range valence
bond state.

To study the AFH model in the entire square lattice we begin by first
tesselating this plane using the cluster of Fig.\ref{fig1} a) as the 
fundamental cell (see Fig. \ref{fig2}). Notice that the centers of the
5-cluster form a new square lattice with lattice spacing $a' =
\sqrt{5} a$.  Given this tesselation we can apply the standard RG
method of replacing clusters of spins by an effective spin
\cite{jullienlibro,jaitisi}. This method has been applied 
for the 1d AFH model by Rabin \cite{rabin} for clusters or blocks
with 3 sites, obtaining a ground state energy with an error of
$12\%$. 
The effective spin of every 3-block in 1d  has spin 1/2.
In our case, as we have discussed above, the effective spin of
the 5-blocks have spin 3S and the energy per block equal to $e_0
= - J S( 4S +1)$. The effective spins $S^\prime=3S$ interact by
means of an effective Hamitonian which to first order in
perturbation theory can be derived if we know the renormalization
of the spin operators ${\bf S}_{\alpha} \rightarrow \xi_{\alpha} 
{\bf S}^\prime ,
\alpha =  0, 1, \dots, 4$.


 The {\em renormalization spin
factor}  $\xi_{\alpha}$ can be shown to be given by the sum
$\xi_{\alpha}=\case{1}{3S} \; \sum_{m_0,m_1,\ldots,m_4} m_{\alpha}
(C^{3S}_{m_0,m_1,\ldots,m_4})^2$ subject to the constraint 
$\sum_{\alpha=0}^4 m_{\alpha}=3S$. 
$ C^{3S}_{m_0,m_1,...}$ is the CG coefficient which
describes the ground state of spin 3S in terms of the 5 original
spins S, whose expression is a product of 4 standard CG
coefficients.
The $\xi_{\alpha}$ satisfy the {\em sum rule}
$\sum_{\alpha=0}^4 \xi_{\alpha}=1$. We arrive at the following result,


\begin{eqnarray} \xi_{\alpha}(S)  = {1\over3S} {6S+1\over 8S+1}
{[(2S)!]^5\over[(8S)!]^2}\cr \times \sum_{m_1,\ldots,m_4} \;  
m_{\alpha}
{(4S-\sum_1^4 m_i)! \; [(4S+\sum_1^4 m_i)!]^2 \over \prod_1^4 (S-m_i)!
\; (S+m_i)! \; [-2S+\sum_1^4 m_i]!}  \label{3}
\end{eqnarray}

\noindent where if $\alpha=0$ then $m_0=3S-\sum_1^4 m_i$. It follows
that the renormalization factors for the four external spins in the
5-block are all equal $\xi_1=\xi_2=\xi_3=\xi_4\equiv \xi (S)$, while
that of the central spin $\xi_0$ is determined by the sum rule.
Amazingly enough the sum (\ref{3}) can be performed in a close manner
yielding,

\begin{equation} \xi (S) = {1\over3} \; {S+\case{1}{4}\over
S+\case{1}{3}} \label{4} \end{equation}

\noindent 
For spin $S=\case{1}{2}$ one obtains $\xi (\case{1}{2})=\case{3}{10}$.
Moreover, Eq. (\ref{4}) 
 correctly reproduces the classical limit $\lim_{S\rightarrow \infty}
\xi(S)=\case{1}{3}$ (recall $S=S^{\text{old}}=\case{1}{3}
S^\prime=\xi_{\text{cl}} S^\prime$). Notice also that the value for
$S=\case{1}{2}$ is already close to the classical value.

The RG-equations for the spin operators $\bf{S}_i$ $i=1,2,3,4$ allows
us to compute the renormalized Hamiltonian $H^\prime$ which turns out
to be of the same form as the original AFH Hamiltonian. In
fact, we arrive at the following RG-equations,

\begin{mathletters} \label{5} \begin{eqnarray} H^\prime(N,S,J) = 
 -J S (4S+1){N\over5} \cr  +
H(\case{N}{5},3 S,3\xi^2(S) J)  \label{5a}
\end{eqnarray} \begin{equation} N^\prime = {N\over5},\; S^\prime = 3
S, \; J^\prime = 3 \xi^2(S) J \label{5b} \end{equation}
\end{mathletters}

\noindent where the first contribution in Eq. (\ref{5a}) comes from
the energy of the blocks. As $3 \xi^2(S)<1$, the flow equation
(\ref{5b}) implies that the coupling constant flows to zero $J^{(n)}
\stackrel{n\rightarrow \infty}{\rightarrow} 0$  which means that the
AFH model remains {\em massless} for arbitrary value of the spin $S$.
This fact allows us to compute the density of energy $e_{\infty}(S)$
(per site) as  the following series,

\begin{mathletters} \label{6} \begin{eqnarray} e_{\infty}(S) =
-\case{1}{5} \sum_{n=0}^{\infty} \case{1}{5^n}  J^{(n)} S^{(n)} (4
\times 3^n S + 1) \label{6a} \end{eqnarray} \begin{equation}
 S^{(n+1)} = 3 S^{(n)}, \; J^{(n+1)} = 3 \xi^2(S^{(n)}) J^{(n)}
\label{6b} \end{equation} \end{mathletters}


Using eqs. (4) and (6) we can compute the ground state energy of
our  variational RG state for any value of the spin S. In
particular for S=1/2 we get the value $e_{\infty}$ = -0.5464. This value
has to be compared with  the ``exact'' numerical result
-0.6692, which is obtained using Green-function Monte Carlo
methods \cite{carlson}. We observe two facts: i) the error of the
computation is $18\%$, which is surprisingly bigger than in 1d
where \cite{rabin} it amounts to a $12\%$  and ii) the result is
closer to the Neel state energy than to the exact ground state
energy of the AFH model. This last observation is confirmed by  
the semiclassical expansion of $e_{\infty}$ in the spin S, which
we  compare  with the standard formulas of Anderson and Kubo
obtained using the spin wave methods \cite{swt},

\begin{eqnarray} & e_{\infty} = - 2 S ( S +  \frac{0.2545}{S} + \cdots
) & \label{6c} \\ & e_{\infty}^{sw} = -2 S ( S + 0.158 +
\frac{0.0062}{S} +\cdots ) & \nonumber \end{eqnarray}

\noindent 
Observe that the lowest order correction to the Neel energy is
absent in our case.


\noindent Yet another explanation for the difference between our 
energy and the numerical value is that the rotating-valence-bond
state as depicted in Fig.\ref{fig2} does not have parallel adjacent bonds,
which excludes the resonance among themselves.

In order to have a better insight into the physics of the model it is
convenient to  compute the staggered magnetization  $M\equiv \langle
\case{1}{N} \sum_j (-1)^j S^z_j \rangle$.
We have been able to  obtain a
closed formula for arbitrary spin $S$ which is capable of analytical
study. To this purpose, we use the RG-equality for V.E.V.  $\langle
\psi_0| {\cal O} |\psi_0 \rangle= \langle \psi^\prime_0| {\cal
O}^\prime |\psi^\prime_0 \rangle$ for renormalized observables ${\cal
O}^\prime$ in the ground state and divide the sum in $M$ into 5-block
contributions. With the help of the renormalization spin factors we
arrive at the RG-equation for the staggered magnetization,

\begin{equation} M_N (S) = {8 \xi (S) - 1\over5} M_{N/5} (3 S)
\label{7} \end{equation}

\noindent The explicit knowledge of $\xi (S)$ (\ref{3}) allows us to
solve this  RG-equation for the staggered magnetization in the
thermodynamic limit  $N\rightarrow \infty$.  In fact, as we know by
now that the Hamiltonian renormalizes to its classical limit, we have 
$\lim_{S\rightarrow \infty} M(S)=S$. Defining $M(S) \equiv S f(S)$,
Eq. (\ref{7}) amounts to solving the equation $f(S)={S+1/5\over S+1/3}
f(3 S)$ subject to the boundary condition $f(\infty)=1$. Thus, we
obtain the following formula for the staggered magnetization for 
arbitrary spin,

\begin{equation} M(S) = S \; \prod_{n=0}^{\infty} {S+\case{1}{5} 3^{-n}
\over S+\case{1}{3} 3^{-n}} \label{8} \end{equation}

\noindent This is a nice formula in several regards. For spin
$S=\case{1}{2}$ we get $M(\case{1}{2})=0.373$ to be compared with
$0.34\pm0.01$ obtained with Green-function Monte Carlo methods
\cite{carlson} and Variational Monte Carlo plus Lanczos algorithm
\cite{hebb-rice}. It amounts to a $7\%$ error.
Other approximate methods
employed so far lead to values of $M(\case{1}{2})$ such as,
 e.g.,  spin
wave theory plus $1/S$-expansion gives \cite{swt} 0.303, spin wave
theory  plus perturbation theory gives \cite{huse} 0.313, etc.
Our value is close to the one found\cite{parrinello}
 with pertubation theory 
around the Ising model to order 4 which is 0.371.
 Another interesting feature of our formula
(\ref{8}) is that it allows us to make a $1/S$-expansion yielding the
result,

\begin{equation} M(S) = S - 0.2 + 0.06 {1\over S} + O(1/S^2)
\label{9} \end{equation}

\noindent which is to be compared with the spin wave theory result
$M(S)=S - 0.198 + O(1/S^2)$ showing excellent agreement for the first
two terms while discrepancies start from the term $1/S$ onwards.


The above  results confirm the picture given at the beginning of
this letter that the rotating-valence-bond state represents a
suggestive proposal to combine the valence-bond and
antiferromagnetic scenarios, but that  resonance phenomena must
be taken into account before claiming that rotation effects take
places among the bonds of the 2d-AFH model.
As was mentioned above, the rotating-valence-bond state so far
introduced in this letter does not allow for resonace phenomena.
However, we can imagine that a bond moves its center of rotation
in the direction of a neighbour bond by hopping (see Fig.\ref{fig3}).
In this case expect that the two bonds will form a resonant pair.
Conversely, we can imagine that a pair of resonant bonds may break
into two rotating neighbour bonds as depicted in Fig.\ref{fig3}.
These intuitive considerations suggest that a 
resonating-rotating-valence-bond scenario may give a satisfactory
picture of the ground state of the 2d AFH model.

\bigskip G.S.  gratefully acknowledges the hospitality of the
Theoretical Physics Institute  at ETH (specially T.M. Rice) while part
of this work was carried out.

Work partially supported in part by 
the Swiss National Science Foundation and by the 
Spanish Fund DGICYT, Ref. PR95-284  (G.S.) and by
CICYT under  contract AEN93-0776
(M.A.M.-D.) .


\vspace{-10 pt}

\begin{figure} \caption{  a) The antiferromagnetic 5-block state.
b) Formation of a rotating-valence-bond state upon applying the $H_{xy}$ part 
of the Hamiltonian to the AF 5-block state.
\label{fig1} } \end{figure}

\begin{figure} \caption{ The two-dimensional square
lattice tesselated by the 5-block.
Dahed lines are nearest-neighbours in the 
renormalized lattice. \label{fig2} } \end{figure}

\begin{figure} \caption{  a) Two rotating bonds pinned 
at the rotation centers $A$ and $C$
respectively.
b) Resonance mechanism for the rotating-valence-bond states. 
Here the rotating bond at $A$
hops to the $B$ site in two different rotating ways.
\label{fig3} } \end{figure}

\end{document}